%% file: main.tex
\documentclass[conference]{IEEEtran}
\ifCLASSOPTIONcompsoc
  \usepackage[nocompress]{cite}
\else
  \usepackage{cite}
\fi
\ifCLASSINFOpdf
\else
\fi
\usepackage{balance}  
\usepackage{graphics} 
\usepackage{txfonts}
\usepackage{times}    
\usepackage[pdftex]{hyperref}
\usepackage{color}
\usepackage{textcomp}
\usepackage{booktabs}
\usepackage{ccicons}

\usepackage{cite}
\usepackage{url}
\usepackage{fancybox}
\usepackage{multirow}
\usepackage{flushend}
\usepackage{booktabs}
\usepackage{tabularx}
\usepackage{comment}
\usepackage{array}
\usepackage[flushleft]{threeparttable}
\usepackage{mdframed}
\graphicspath{{}{images/}{dia/}}
\DeclareGraphicsExtensions{.pdf,.png}

\usepackage{listings}
\usepackage{courier}
\usepackage{hyperref}

\usepackage{xpatch}

\xpatchcmd{\refstepcounter}{%
  \stepcounter{#1}%
}{%
  \stepcounter{#1}%
  \xdef\scr{\number\value{#1}}%
}{\typeout{success}}{\typeout{failure}}

\newcounter{o}
\usepackage{tikz}
\definecolor{1c1}{RGB}{188,162,6}
\definecolor{1c2}{RGB}{137,129,80}
\definecolor{1c3}{RGB}{239,167,31}
\definecolor{1c4}{RGB}{88,194,241}
\definecolor{1c5}{RGB}{6,180,188}

\tikzset{mynode/.style={draw=white,solid,circle,fill=green,inner sep=1pt, thick,
text=black}}
\tikzset{arrow line/.style={dashed, line width= 2.5pt, color=#1}}

\def\bf{\textbf}

\def\fig {Figure~}

\def\tbl {Table~}

\def\it{\textit}

\def\tt{\mct}

\newcommand{\mct}[1]{{\footnotesize {\texttt {#1}}}}

\usepackage{paralist}

\usepackage[small,bf]{caption}
\usepackage{tikz}

\lstdefinestyle{inlinecode}{basicstyle={\ttfamily\scriptsize\bfseries}}

\newcommand{\urls}[1]{{\scriptsize\url{#1}}}
\usepackage{tcolorbox}

\usepackage{paralist}
\usepackage[outercaption]{sidecap}
\usepackage [autostyle, english = american]{csquotes}
\MakeOuterQuote{"}
\newcounter{scn}
\usepackage[shortlabels]{enumitem}
\usepackage{tikz}
\usepackage{styles/pgf-pie}
\usetikzlibrary{positioning,shadows}
\usepackage{balance}

\newif\ifpienumberinlegend
\pgfkeys{/number in legend/.code=
    \expandafter\let\expandafter\ifpienumberinlegend
    \csname if#1\endcsname
    \ifpienumberinlegend

    \def\beforenumber##1\afternumber{}%
    \fi,
    /number in legend/.default=true
}
\definecolor{1c1}{RGB}{188,162,6}
\definecolor{1c2}{RGB}{137,129,80}
\definecolor{1c3}{RGB}{239,167,31}
\definecolor{1c4}{RGB}{88,194,241}
\definecolor{1c5}{RGB}{6,180,188}

\tikzset{mynode/.style={draw=white,solid,circle,fill=green,inner sep=1pt, thick,
text=black}}
\tikzset{arrow line/.style={dashed, line width= 2.5pt, color=#1}}

\definecolor{1c1}{RGB}{188,162,6}
\definecolor{1c2}{RGB}{137,129,80}
\definecolor{1c3}{RGB}{239,167,31}
\definecolor{1c4}{RGB}{88,194,241}
\definecolor{1c5}{RGB}{6,180,188}

\tikzset{mynode/.style={draw=white,solid,circle,fill=green,inner sep=1pt, thick,
text=black}}
\tikzset{arrow line/.style={dashed, line width= 2.5pt, color=#1}}

\usepackage{bchart}

\usepackage{xcolor}
\definecolor{ao(english)}{rgb}{0.0, 0.5, 0.0}

\def\test#1{%
\ifnum0#1>0
      #1
    \fi
}

\newcommand{\sevenbars}[7]{
{{\color{black}\rule{#1pt}{4pt}} \test{#1}}
{{\color{ao(english)}\rule{#2pt}{4pt}} \test{#2}}
{{\color{magenta}\rule{#3pt}{4pt}} \test{#3}}
{{\color{cyan}\rule{#4pt}{4pt}} \test{#4}}
{{\color{red}\rule{#5pt}{4pt}} \test{#5}}
{{\color{yellow}\rule{#6pt}{4pt}} \test{#6}}
{{\color{orange}\rule{#7pt}{4pt}} \test{#7}}
}
\usepackage{longtable}

\usepackage{bchart}

\usepackage{pgfplots}
\pgfplotsset{compat = newest}

\usepackage{appendix}

\begin{document}
\title{A Large-Scale Empirical Study of COVID-19 Contact Tracing Mobile App Reviews}
\author{
\IEEEauthorblockN{
Sifat Ishmam Parisa\IEEEauthorrefmark{1}, Md Awsaf Alam Anindya\IEEEauthorrefmark{1}, Anindya Iqbal\IEEEauthorrefmark{1}, Gias Uddin\IEEEauthorrefmark{2}
}
\IEEEauthorblockA{\IEEEauthorrefmark{1} Bangladesh University of Engineering \& Technology}
\IEEEauthorblockA{\IEEEauthorrefmark{2} University of Calgary}
\IEEEauthorblockA{1505016.sip@ugrad.cse.buet.ac.bd, 1505114.maaa@ugrad.cse.buet.ac.bd , anindya@cse.buet.ac.bd, gias.uddin@ucalgary.ca}
}

\IEEEtitleabstractindextext{%
\begin{abstract}
Since the beginning of 2020, the novel coronavirus has begun to sweep across the globe. While the recent development of vaccines offers hope, the mutations of the coronavirus have enforced 
countries around the world to adopt and enforce strict safety requirements like wearing a mask. 
Given the prevalence of smartphones everywhere, many countries across continents also developed COVID-19 contract tracing apps that users can install to get a warning of potential contacts with infected people. Unlike regular apps that undergo detailed requirement analysis, carefully designed development, rigorous testing, contact tracing apps were deployed after rapid development. Therefore such apps may not reach expectations for all end users. Users share their opinions and experience of the usage of the apps in the app store. This paper aims to understand the types of topics users discuss in the reviews of the COVID-19 contact tracing apps across the continents by analyzing the app reviews. This insight can be useful to learn about the specific problems the users face, the concerns they have about the apps, and the positive/negative experience they share about the design and usage of the apps. We collected all the reviews of 35 COVID-19 contact tracing apps developed by 34 countries across the globe. We group the app reviews into the following geographical regions: Asia, Europe, North America, Latin America, Africa, Middle East, and Australasia (Australia and NZ). The purpose is to ensure that we can capture region-specific diversity and specificity during the analysis of the app reviews. We run topic modeling on the app reviews of each region. We analyze the produced topics and their evolution over time by categorizing them into hierarchies and computing the ratings of reviews related to the topics. While privacy could be a concern with such apps, we only find the privacy-related topics in Australasia, North America, and Middle East. Topics related to usability and performance of the apps are prevalent across all regions. Users frequently complained about the lack of features and negative/hung user interfaces in the apps, as well as the negative impact of such apps on their mobile batteries. Despite the negative experiences, we also find that many users praised the apps because they helped them stay aware of the potential danger of getting infected. The finding of this study is expected to help the app developers utilize their resources in addressing the reported issues in a prioritized way.
\end{abstract}

}

%


\maketitle

\IEEEdisplaynontitleabstractindextext

%
\IEEEpeerreviewmaketitle

\input{introduction}
\input{background}
\input{study-setup.tex}

\input{empirical-study.tex}
\input{discussions.tex}
\input{threats}
\input{conclusion}

\begin{small}
\bibliographystyle{abbrv}
\bibliography{consolidated}
\end{small}

\end{document}

%% file: introduction.tex
\section{Introduction}\label{sec:introduction}
The global health crisis of 2020-21 is the ongoing COVID-19 pandemic, caused by  Severe Acute Respiratory Syndrome Coronavirus 2 (SARS-CoV-2) \cite{velavan2020covid}. The worldwide outbreak compelled governments to adopt countermeasures such as lockdown, social-distancing, wearing masks, self-isolation, work-from-home, and contact tracing \cite{contacttracing}. Conventional contact tracing is a manual procedure of approximating the risk score of contacts of the infected individuals conducted by health officials. However, with the high number of cases, the traditional contact tracing approach for COVID-19 is complicated, time-consuming, and not entirely effective. Hence, authorities and stakeholders have relied on data analytics and digital technologies to automate the contact tracing process. Consequently, several smartphone apps for contact tracing have been developed \cite{o2020flood}. Although vaccination has started to be administered in developed countries, it has been predicted that it will take more than the whole 2021 to bring required number of people across the globe to achieve hard immunity. Hence, to combat the ongoing second wave and for a long time afterwards, the role of Contact Tracing Applications remains highly significant.
\newline
Contact tracing apps notify a person who has been in the proximity of Covid infected patients with the help of smartphones' built-in services such as Bluetooth interface and location service. These apps differ from regular apps in the following ways. Regular application development is expected to follow all standard software development steps, i.e., planning, requirement analysis and documentation, functionality design, programming, comprehensive testing, deployment, implementation, and support. On the contrary, contact tracing apps were developed as an emergency response tool to control the virus spread. The time constraint did not allow careful requirement planning and feature design. It is also highly likely that the design and testing phase did not take all requirements into account. As a result, using the apps was problematic for the majority of users, as found in their app reviews.
\newline
Users can submit their opinions regarding app usage experience by star ratings and text reviews in online application distribution platforms like Google Play Store and App Store. As presented in empirical studies, \cite{msrforappstores}, \cite{softwarerequirementsevolution}, \cite{appstoreempiricalstudy}, and \cite{preliminaryanalysis}, these reviews reveal user experience and requirements, bug reports, and feature requests. Every app development team can get an idea of their own strengths/weaknesses as well as that of similar other applications. Developers can use the data gathered from the reviews as guidelines for updating the application in subsequent releases \cite{mobilere}. It is likely that different development teams are supposed to address particular types of issues (e.g., front-end/UX, authentication/security, etc.) that need specialization to some extent. We have identified the topics of the reviews and then grouped the topics into high-level categories that would help the management assign issues easily to respective development groups. Therefore, we expect the outcome of this study would help the continuous improvement of the contract tracing applications and benefit the community. Moreover, the insights gained will be useful for any future emergency situation where such quick developments will be required. The apps being developed for nationwide vaccination management are examples of such emergency response application development that may benefit from the insight presented by this study.
\newline
To conduct this study, we collected a total of 455640 app reviews of 35 contact tracing apps released by governments/official authorities of 34\footnote{We collected data from 2 contact tracing apps of Malaysia.} countries around 7 continents. Then we distribute the reviews to 7 corpora and apply topic modeling to each corpus independently. We label the topics discovered by the topic model and classify them into  main categories. Then we systematically analyzed the identified topics in the collection of reviews.  

Specifically, we explore the following research questions in this study:
\begin{enumerate}
  \item What are the topics discussed in COVID-19 contact tracing app reviews across the continents?\\
  We identify 31 topics in the app reviews. The topics can be grouped into seven categories - "Usability", "Reliability", "Performance", "Privacy and security", "Configuration", "Feature shortcoming", and "Version". Among the categories, Usability is the most popular one.
  \item How does the popularity of the topics evolve over time?\\
  We investigate the time-specific trends of the reviews and discover that user activity was at its highest level from April to July in 2020. During this period, the frequency of reviews under each category increased up to May and then gradually decreased.
  \item How do the complaints and appraisals of the users vary across the topics?\\
  Based on star reviews, we discover that 88\% of reviews under Usability are user appraisals, while the reviews under the other categories are more about complaints than appraisals.
  \item How do the complaints and appraisals of the users evolve across the continents?\\
  We analyze the ratio of appraisal and complaint for each continent and observe that 88\% of reviews are appreciative in Asia, higher than other continents. Europe, North America, Africa, and Latin America show relatively more appraisals than complaints. In the other continents, the ratio between appraisal and complaint differ marginally.
\end{enumerate}

To the best of our knowledge, this is the first study that focused on understanding the COVID-19 contact tracing app reviews. The contribution of this paper is threefold:
\begin{itemize}
    \item[$\bullet$] We have compiled a dataset of reviews of 35 COVID-19 contact tracing apps and divided them into seven corpus based on continents.
    \item[$\bullet$] We then divide the topics discussed into seven categories.
    \item[$\bullet$] Finally, we study the evolution of the topics across the continents.
\end{itemize}

The remainder of this paper is organized as follows. Section~\ref{sec:related} introduces the background concepts and related works followed by our approach in section~\ref{sec:studysetup}. Section~\ref{sec:empiricalstudy} reports the results of the research questions. Section~\ref{sec:discussion} discusses the findings and their implications. Section~\ref{sec:threats} discusses threats to validity while section~\ref{sec:conclusion} presents conclusive remarks.

%% file: background.tex
\section{Related Work}\label{sec:related}
Due to the huge importance, the Contract Tracing apps have gained attention from researchers right from the beginning. They have conducted multiple studies from different perspectives. Also, other software engineering research works are carried out related to various aspects of Covid 19. In this section, we briefly discuss these prior works. Since this is a topic modeling based study, previous similar studies are also explored.
\subsection{Studies on COVID-19 Contact Tracing App}
The privacy concern of the users was highly pronounced since the inception of contact tracing apps. This was reflected in the academic research also as discuss below. The underlying protocols, evaluation methods, and deployment architecture were also formally studied by the researchers.\\
Vaudenay \cite{vaudenay2020centralized} provides a comprehensive study of the advantages and vulnerabilities of centralized and decentralized contact tracing systems. Redmiles \cite{redmiles2020user} explores user tradeoffs for different contact tracing architectures. Functions, evaluation methods, and the cost assessment of digital and traditional contact tracing methods are discussed in \cite{von2020research}. Simko et al. \cite{simko2020covid} survey public opinion on privacy concerns regarding contact tracing apps. Sun et al. \cite{sun2020vetting} developed a security and privacy tool to examine security issues, privacy leaks, embedded trackers, and malwares in contact tracing apps. Azad et al. \cite{azad2020first} conduct a manual investigation on privacy, security, and permission requirements of contact tracing apps. Li et al. \cite{li2020covid} review the deployment of centralized and decentralized architectures. Reichart et al. \cite{reichertsurvey} present a survey on various automatic contact tracing approaches and classified them on privacy risk score. Tang \cite{tang2020privacy} studies privacy issues for privacy-aware contact tracing solutions and proposes an interdisciplinary research agenda to potential solutions. Cho et al. \cite{cho2020contact} study the privacy issues of contact tracing apps in depth. They propose techniques to develop a contact tracing app aiming to incorporate privacy enhancements for all users. Ahmed et al. \cite{ahmed2020survey} present the key attributes of contact tracing apps, i.e., architecture, data management, privacy, security, contact estimation, and attack vulnerability. They study 15 tracing apps based on different tracing protocols.

\subsection{Software Engineering research related to COVID-19}\label{sec:relatedworkCOVIDSE}
Software engineering research related to COVID-19 is limited. We have identified three recently published/shared 
papers in software engineering that have focused on the pandemic. Two of the papers~\cite{ralph2020pandemic,bao2020does} investigate the impact of productivity and 
wellbeing of developers during the pandemic, while the third paper~\cite{rahman2020exploratory} analyzes the bugs in COVID-related apps. 

The recently published article by Ralph et al.~\cite{ralph2020pandemic} 
surveyed 2,225 developers from 53 different countries 
and asked them questions about how their productivity and wellbeing are affected by COVID-19. They found that in general productivity and 
wellbeing have declined and they are closely related. This means that lack of wellbeing can impact a loss in productivity. They also found heterogeneity 
in the needs of the developers from their organizations, i.e., different people expected and needed different types of support. 
They also find that the pandemic may have affected more severely women, parents, and people with disabilities.   
 
The second paper, still unpublished but generously shared by the authors Bao et al.~\cite{bao2020does} in arXiv, focuses on understanding wellbeing  
and productivity of Baidu employees while working from home during the COVID-19 pandemic. Baidu is a large Chinese software company. 
The paper quantitatively analyzes 4,000 records (e.g., code commits, reviews, etc.) 
of 139 developers' activities of 138 working days. Out of the records, 1,103 records are submitted when the 
the developers worked from home due to the COVID-19 pandemic. Contrary to Ralph et al.~\cite{ralph2020pandemic}, they found that 
work from home has both positive and negative impacts on developers' productivity in terms of different metrics, such as the number of commits.

Rahman and Farhana~\cite{rahman2020exploratory}
studied bugs in open source projects related to COVID-19. They identified 8
categories of bugs in 7 different kinds of projects. 

\subsection{Topic Modeling in Software Engineering Research}
Our motivation to use topic modeling to understand COVID-19 contact tracing app reviews stems from existing research in software engineering that shows that topics generated from textual contents can be good approximation of the underlying \it{themes}~\cite{Chen-SurveyTopicInSE-EMSE2016,Sun-SoftwareMaintenanceHistoryTopic-CIS2015,Sun-ExploreTopicModelSurvey-SNPD2016}. Topic models are used recently to understand software logging~\cite{Li-StudySoftwareLoggingUsingTopic-EMSE2018} and previously 
for diverse other tasks, such as concept and feature location~\cite{Cleary-ConceptLocationTopic-EMSE2009,Poshyvanyk-FeatureLocationTopic-TSE2007}, 
tracability linking (e.g., bug)~\cite{Rao-TraceabilityBugTopic-MSR2011,AsuncionTylor-TopicModelingTraceabilityWithLDA-ICSE2010a}, to understand software and source code history 
evolution~\cite{Hu-EvolutionDynamicTopic-SANER2015,Thomas-SoftwareEvolutionUsingTopic-SCP2014,Thomas-EvolutionSourceCodeHistoryTopic-MSR2011}, to facilitate code search by categorizing software~\cite{Tian-SoftwareCategorizeTopic-MSR2009}, to understand the challenges of blockchain development~\cite{SOBCSAlahi}, to understand the evolution of new languages ~\cite{3NewLang},
to refactor software code base~\cite{Bavota-RefactoringTopic-TSE2014}, as well as to explain software defect~\cite{Chen-SoftwareDefectTopic-MSR2012} and various 
software maintenance tasks~\cite{Sun-SoftwareMaintenanceTopic-IST2015,Sun-SoftwareMaintenanceHistoryTopic-CIS2015}. In between studies also focused on the effective 
use and parameterization of topic modeling for software engineering tasks~\cite{Panichella-EffectivelyUseTopic-ICSE2013,Panichella-ParameterizationTopic-SANER2016}. In this 
paper, we adhere to the recommended parametertization of topic modeling. 
The Stack Overflow (SO) posts are subject to a number of recent papers to study various aspects of software development using topic modeling, such as what developers are discussing in general~\cite{Barua-StackoverflowTopics-ESE2012},
or about a particular aspect, e.g., concurrency~\cite{Ahmed-ConcurrencyTopic-ESEM2018}, big data~\cite{Bagherzadeh2019}, chatbot development~\cite{abdellatifchallenges}. 
We are aware of no previous 
research which specifically focused on understanding the COVID-19 contact tracing app reviews.  
 

%% file: study-setup.tex
\section{Data Collection and Topic Modeling}\label{sec:studysetup}

In this section, we discuss the data collection and preprocessing steps and also the methodology adopted for topic modeling.
\subsection{Data Collection and Preprocessing}
We download app review data from Google Play Store through an open-source scraping tool google-play-scraper\footnote{\url{https://github.com/JoMingyu/google-play-scraper}}. It retrieves the reviews with the additional information, i.e., a unique review ID, user name, star rating, timestamp, count of thumbs up of the review, and app version corresponding to the review. Our initial dataset contains 4,55,640 reviews in total from 35 apps.
\newline
The raw data is very noisy, so we perform the following preprocessing steps.
\begin{itemize}
    \item First, we discard the URL and emoticons from the reviews. We use an open-source tool called demoji\footnote{\url{https://github.com/bsolomon1124/demoji}} to detect and remove emoticons.
    \item Then we use Googletrans\footnote{\url{https://github.com/ssut/py-googletrans}}, a python google translate API, to detect a particular review's source language. We only considered reviews that are in English.
    \item English words do not have the same letter three consecutive times. We modify words containing the same letter continuously for more than three times to two occurrences of that letter. This step, for example, converts the word "goooood" to "good".
    \item We inspect the most frequent numerical terms in the reviews and select the relevant words to include and discard all other numbers. An example of a selected numerical term is "109" representing the helpline number of Qatar's contact tracing app.
    \item Smartphone model name with number is often found in the reviews. We search for such device identity-specific terms and collapse them into the term "phone". Hence, when a review contains the words "Samsung A40", this step transforms the words to "phone".
    \item We use a list of stopwords containing common English words that contribute nothing to the topics. Our stopword list comprises the default list of stopwords in Natural Language Toolkit, NLTK \cite{bird2009natural}. The list also contains specific stopwords such as the name of an application, country name, country code, etc. that we use independently for each application. Terms frequently used in application reviews are also included in the stopword list. Examples of such terms are "app", "please", "fix", etc.
    \item Stemming reduces inflectional and derivationally related forms of a word to a common base form. We use snowball stemmer by NLTK. For example, stemming maps the words "permission", "permissions", and "permissible" to "permiss".
\end{itemize}
\subsection{Topic Modeling}
Topic modeling is a statistical approach for discovering abstract topics from a collection of discrete data. One of the most popular tools for topic modeling is Latent Dirichlet Allocation (LDA) \cite{blei2003latent}. LDA is a generative probabilistic model for text documents. The assumption is that documents are generated according to latent topics where each topic is a multinomial distribution over the vocabulary. LDA assumes the following steps for generating each document. For each word in the document, LDA samples a topic from a multinomial distribution of a previously sampled Dirichlet random variable. Then words are sampled from a multinomial distribution conditioned on the selected topic. Given a document, the posterior distribution of LDA model parameters and hidden variables can reveal latent semantics in the corpus to classify documents of different topics. However, the posterior distribution is computationally intractable, so inference algorithms such as variational approximation, Markov Chain, Monte Carlo, etc. are used. Hoffman et al. \cite{hoffman2010online} proposed an online variational inference algorithm for LDA. It is based on online stochastic optimization, does not need to store the collection of documents, and converges faster than batch algorithms on large datasets. We use the open-source NLP framework Gensim \cite{rehurek_lrec} for topic modeling. The core estimation for LDA in Gensim is the online variational inference algorithm.
\newline
We group the app reviews into seven corpora according to the continents. Then, we apply LDA to each corpus individually. We experiment by varying the number of topics, parts of speech selection, and dictionary filtering. The number of topics is a model parameter that plays a vital role in performance improvement. The coherence score \cite{newman2010automatic} is an appropriate metric for evaluating the interpretability of the discovered topics. We use Gensim module CoherenceModel that implements the four stage topic coherence pipeline proposed by R{\"o}der et al.~\cite{roder2015exploring} to compute the topic coherence for a specific value of the number of topics. We experiment for different values to maximize the coherence value. Furthermore, we manually check the top 15 reviews of a topic based on the probability of being assigned to that topic. Relying on the coherence value and manual examination, we finalize the number of topics for each corpus. After the preprocessing and topic modeling pipeline, we have 370,524 labelled reviews.

\begin{table*}[t]
  \centering
   \caption{Distribution (frequency) of topics in our dataset per category. Each colored bar denotes a continent
   (Black = Europe, Green = North America, Magenta = Australasia, Blue = Asia,  Red= Middle East, Yellow = Africa, Orange = Latin America)}
    \resizebox{\textwidth}{!}{%
    \begin{tabular}{lr}
    \toprule{}
    
   
    \textbf{Topics} & \textbf{Frequency of Topics Observed}\\
    \midrule
    \textbf{Usability} &  \\
    Satisfactory UI &  \sevenbars{15}{16}{0}{0}{0}{0}{0} \\
    UI suggestions &  \sevenbars{8}{0}{3}{2}{0}{12}{0} \\
    Input suggestions &  \sevenbars{0}{0}{0}{1}{0}{0}{0} \\
    UI issues &  \sevenbars{0}{0}{0}{1}{0}{0}{40} \\
    Unsatisfactory usage &  \sevenbars{5}{0}{0}{0}{3}{0}{0} \\
    Uncertain feasibility &  \sevenbars{0}{11}{0}{0}{0}{0}{0} \\
    Customer support issue &  \sevenbars{0}{0}{0}{0}{4}{0}{0} \\
    Mixed polarity in UX &  \sevenbars{0}{0}{4}{2}{0}{0}{0} \\
    UX evolution &  \sevenbars{0}{0}{4}{0}{0}{0}{0} \\
    Positive UX &  \sevenbars{6}{17}{19}{77}{31}{35}{11} \\
    Download and installation UX &  \sevenbars{0}{0}{9}{0}{0}{0}{0} \\
    \midrule
    \textbf{Reliability} &  \\
    Unsatisfactory results &  \sevenbars{9}{0}{0}{4}{5}{0}{19} \\
    Contact trace functionality issue &  \sevenbars{0}{11}{0}{1}{0}{0}{0} \\
    App crash &  \sevenbars{0}{0}{4}{0}{14}{0}{0} \\
    Phone number format handling &  \sevenbars{0}{0}{14}{0}{0}{0}{0} \\
    Results issue &  \sevenbars{0}{0}{0}{0}{0}{20}{0} \\
    \midrule
    \textbf{Performance} &  \\
    Battery usage &  \sevenbars{6}{0}{2}{0}{6}{0}{0} \\
    Data load issue &  \sevenbars{0}{0}{0}{0}{0}{0}{0} \\
    Response error &  \sevenbars{0}{0}{0}{1}{0}{0}{0} \\
    \midrule
    \textbf{Privacy and security} &  \\
    QR code &  \sevenbars{9}{0}{5}{0}{0}{0}{0} \\
    Controlled access &  \sevenbars{10}{0}{0}{0}{3}{0}{0} \\
    Login error &  \sevenbars{0}{0}{7}{2}{0}{0}{0} \\
    Verification code issue &  \sevenbars{0}{0}{6}{0}{6}{0}{15} \\
    Privacy concern &  \sevenbars{0}{9}{3}{0}{4}{0}{0} \\
    \midrule
    \textbf{Configuration} &  \\
    Impact on phone components &  \sevenbars{7}{20}{2}{2}{4}{31}{0} \\
    Bluetooth interference &  \sevenbars{0}{0}{5}{0}{0}{0}{0} \\
    Location issues &  \sevenbars{0}{0}{2}{0}{0}{0}{0} \\
    \midrule
    \textbf{Feature shortcoming} &  \\
    Location estimation &  \sevenbars{5}{0}{0}{0}{0}{0}{0} \\
    Language issue &  \sevenbars{0}{0}{0}{0}{0}{0}{13} \\
    \midrule
    \textbf{Version} &  \\
    Version Update &  \sevenbars{3}{0}{0}{0}{9}{0}{0} \\
    Compatibility &  \sevenbars{8}{13}{4}{2}{6}{0}{0} \\

    \bottomrule
    \end{tabular}%
   }
  \label{tab:challengespertopic}%
\end{table*}%

\begin{table*}[hbt!]
    \centering
    \small
    \begin{tabular}{ |p{0.15\linewidth} |p{0.7\linewidth} |}
    \hline
Topic & Example(s) \\ \hline
Impact on phone components & I didn't open the app for a while and I have noticed that the database got updated in the moment I opened, this means is not correctly working in the background. A suggestion to prevent the app from going to sleep could be to implement "ongoing notifications", this is done by Google for the weather or Fitbit for MobileTrack. Another feature I've noticed from such apps that need to always run in the background is to disable battery optimisation. Good luck. \\ \hline 
Bluetooth interference & Since installing this app, every other Bluetooth use for my phone is no longer working. I can't listen to music on Bluetooth speaker without it constantly cutting out and I can't use hands free in my car. \\ \hline 
Location issues & Privacy policy says "No location data (data that could be used to track your movements) will be collected at any time." Then the app asks for permission to use location data. Not sure why it needs this permission it if does not use the data? \\ \hline 
Location estimation & Using the first half of your postcode to work out your risk level is rubbish. My postcode area is mostly in Cumbria and Lancashire, with only a little bit of Yorkshire (where I live). It says 'my area or a neighbouring area' is in lockdown. I am not in lockdown. The information provided is not sufficient to be useful - it needs to say 'you are in lockdown' or 'you are not in lockdown'. Saying 'you might possibly be in lockdown' is a total waste of time... \\ \hline 
Language issue & Why is the app description in English if the content isn't? \\ \hline
Data load issue & It takes too long time to update, some time not working properly \\ \hline 
Response error & After entering my details the submit button is not taking to the next activity. And when i open it again it is asking me to enter details again.. \\ \hline 
QR code & Useless. Installed ok but every place I visit it cant scan and says QR code not in correct format...or place doesn't have QR codes yet ...like Mitre 10! \\ \hline 
Controlled access & Pointless - it's the official launch date, but still can't access it as it's only for the IoW and people invited to trial it. Apparently loads of people signed up today - I've no idea how! Bloody useless. \\ \hline 
Privacy concern & There is no way this app doesn't collect your location. They know the IP address, device ID and nearby devices. I will not use this app due to privacy concerns. This app will not help protect my life more than a drops of hand sanatizer. \\ \hline 
Unsatisfactory results & Doesn't show correct position of covid-19 patients, a family tested positive within 500m of my range and it shows no covid-19 patients within 5kms of my range. The no. of covid-19 cases shown near your area isn't also accurate. \\ \hline 
Contact trace functionality issue & Wasnt really helpful. I got notified that I've been exposed in the last 14 days but I dont know if thats someone in a car beside me, when I was in contact with them, basically any information to tell me if I'm actually at risk or not. It would be more helpful if it said the day you were in contact with the person and the time so you can figure out if you're actually at risk \\ \hline 
Results issue & The app is a great idea but this is South Africa, the majority of people who tested positive will not register themselves. This app should have been connected with the databases of all labs who have the contact details of people who tested positive. Obviously, their I formation will be kept confidential, people who downloaded the app will have the advantage of more accurate warnings. This app is currently a false security. \\ \hline 
UI suggestions & Please add covid positive patient location on map view in application Also show my range area circle in map view thank you \\ \hline 
Input suggestions & In proffesion column you didn't add bank employee who are highly exposed during the lockdown. Kindly add bank employee in the proffesion column. \\ \hline 
Uncertain feasibility & I use it I wish more people would use it it's safe it doesn't collect your information and if 60\% of Canadians actually use this app it would make a huge difference right now I think it's about 3\%, or 10\%, I don't know it's not nearly enough but come on people we're better than this \\ \hline 
Customer support issue & It worked smoothly since last week. Now always asking to varify mobile and later mobile number blocked after reach out limit and asking to call 109. But they says wait for 2 days or use another number. How may time we can change number. Is it not possible to clear this issue from customer care center \\ \hline 
Mixed polarity in UX & This is a third class aap not a suitable for a war of covide19, Well done work on this app \\ \hline 
UX evolution & First time I used the app was not working, now that we are lvl3 again work much better. \\ \hline 
Download and installation UX & I have tried to download the app for 3 nights and waited up to an hour for the download to start. I have given up trying and if most people experience the same trouble, then the government is not going to get the numbers required. \\ \hline 
Version Update & I've had the app for almost three weeks, but the first page has been saying "13 of 14 days active" for the last few days. Reports in the press two days ago said this was a glitch and there was an update version 1.0.4 for Android that would fix it. I checked the RKI website but there is no such update. \\ \hline 
Compatibility & Does not work unless you have a phone running at least Android 6 or IOS 13.5. So iPhone 5 won't work.. \\ \hline 
 
    \end{tabular}
    \caption{Representative example app-reviews of 22 topics}
    \label{tab:example}
\end{table*}

%% file: empirical-study.tex
\section{Empirical Study}\label{sec:empiricalstudy}
In this research, we want to systematically study the issues reported in the user reviews of different contact tracing apps across different continents. We explore the following research questions to develop an insight into the user feedback:
\begin{enumerate}
  \item What are the topics discussed in COVID-19 contact tracing app reviews across the continents?
  \item How does the popularity of the topics evolve over time?
  \item How do the complaints and appraisals of the users vary across the topics?
  \item How do the complaints and appraisals of the users evolve across the continents?
\end{enumerate}

\subsection{Topics discussed in COVID-19 contact tracing app reviews (RQ1)}
\subsubsection{Motivation} 
A major benefit we may receive by analyzing the app reviews is to have insight about user requirement of new features, bug reports, user experience(UX) and security issues, etc. The categorization should be helpful to assign development teams for addressing those. Due to the substantial difference in the nature of development of different software components, e.g., UI, security mechanisms, etc., it would expedite the new release if issues under a topic are addressed by the team specialized in it.

\subsubsection{Approach} We determine 31 topics from 7 continents with overlapping topics among continents. Individually, the topic count is 12 in Europe, 7 in North America, 16 in Australasia, 12 in Asia, 12 in Middle-East, 4 in Africa, and 5 in Latin America. We assign the dominant topic to a review. The dominant topic is the topic with the maximum probability in that review's topic distribution. We categorize the topics into seven software engineering specific categories, i.e., "Usability", "Reliability", "Performance", "Privacy and security", "Configuration", "Feature shortcoming", and "Version".
\subsubsection{Results}

\tbl\ref{tab:challengespertopic} displays the topics determined by our approach. The first column holds the topics grouped into categories. The next column represents the percentage of topics across different continents. Each color denotes a continent, followed by the continent's topic frequency.
\newline
UX related reviews are noticed across all the continents, where the majority of them are satisfactory. 
Users also propose suggestions for the improvement of the app's user interface(UI). In Latin America, reviews under Usability mostly reveal UI issues such as complicated UI of the app. In Asia, maximum reviews are from the contact tracing app of India. Users are satisfied with the app's overall functionality; hence, 77\% of the reviews show the feeling of positive UX. Users in North America, Australasia, and Middle East are concerned about the app's privacy threat. Security issues involving authorization and authentication also exist in reviews across all the continents. Another common pattern in reviews across most continents is the app's compatibility with devices and the operating system. In some cases, the app cannot operate on older devices. Also, the app does not work on the operating system of certain versions. In general, the apps of all the continents have a major drawback, i.e., they drain most of the battery power in less than acceptable time. Users also express concern about the app's impact on different components of the phone, e.g., battery, Bluetooth, GPS, which indicates that some performance optimization techniques may not be properly implemented in the quick development period. \\
\tbl\ref{tab:example} shows a representative example of each topic.

\subsection{Evolution of the popularity of topics over time(RQ2)}
\subsubsection{Motivation}
There is a difference in culture and regulations of the countries of different continents. For example, the privacy concern of the app users in Europe or North America may vary from that of underdeveloped countries. The types/configurations of phones used by the mass population depend on the economic condition of the country. The Government of some countries can enforce safety policies in a more strict way than the others where the privacy concern of the citizens is given high importance. The impact of such differences on app users' experience is expected to be revealed from the analysis of reviews. Examining the time-specific trends of the reviews will give an overview of the app's capability to overcome issues. The evolution of reviews across time can show emerging challenges with the app. 

\subsubsection{Approach} In order to examine the time-specific trends of popularity, we compare the absolute impact and relative impact of the topics under different categories. The choice of these metrics is made from the consideration that frequency based metrics only consider the dominant topic with the highest probability and may miss the involvement of other topics. The absolute impact score of a topic in a particular month is the sum of the topic's probabilities for all reviews in that month. Therefore, the absolute impact of a category sums up the absolute impact of all topics under this category. We calculate this metric for all categories in the time period of February to September of 2020. The relative impact score of a topic in a specific month is the corresponding absolute impact scaled by the total reviews in that month. We compute the relative impact of a category by summing up the relative impact of topics under the category. Finally, we plot the relative impact score of different categories between the aforementioned time interval. 
\subsubsection{Results} 
\begin{figure*}[h]
\centering
\begin{tikzpicture}
\begin{axis}[
    scaled ticks=false, tick label style={/pgf/number format/fixed},
    ylabel={Relative Topic Impact},
    x=2cm,
    xmin=0, xmax=7,
    ymin=0, ymax=0.7,
    xtick={0, 1, 2, 3, 4, 5, 6, 7},
    xticklabels={Feb-20, Mar-20, Apr-20, May-20, Jun-20, Jul-20, Aug-20, Sep-20},
    ytick={0,0.1,0.2,0.3,0.4,0.5,0.6,0.7},
    xtick pos=bottom,
    ytick pos=left,
    legend style={at={(axis cs:2,0.17)},anchor=south west},
    ymajorgrids=true,
    grid=none,
]
\addplot[color=black,]
coordinates {(0,0.6250007189810276)(1,0.3609450704953255)(2,0.6178121048825312)(3,0.6462183931561711)(4,0.5931253386373021)(5,0.5964754718041905)(6,0.5048984874880736)(7,0.4477598588630568)};
\addlegendentry{Usability}
\addplot[color=blue,]
coordinates {(0,0.026785721071064472)(1,0.12068979923325855)(2,0.08215578452102497)(3,0.07458776329528419)(4,0.10009895660557112)(5,0.1010341998717912)(6,0.1027592784385684)(7,0.10086970311201313)};
\addlegendentry{Reliability}
\addplot[color=red,]
coordinates {(0,0.14731928519904613)(1,0.060214351656319824)(2,0.04749462782252021)(3,0.046173405210528366)(4,0.04487286551742725)(5,0.04406441174816873)(6,0.054266902182742084)(7,0.0410533291454153)};
\addlegendentry{Performance}
\addplot[color=green,]
coordinates {(0,0.008928573690354824)(1,0.13121004684923146)(2,0.037480716948457726)(3,0.029499626318192963)(4,0.045543398127856884)(5,0.037606129374516746)(6,0.09666645196465604)(7,0.13543699278115687)};
\addlegendentry{Privacy and security}
\addplot[color=orange,]
coordinates {(0,0.008928573690354824)(1,0.10395822764554527)(2,0.03019440278759317)(3,0.026351433795392586)(4,0.032153864095341385)(5,0.03303412670332866)(6,0.060138516748061525)(7,0.07191813722691877)};
\addlegendentry{Configuration}
\addplot[color=yellow,]
coordinates {(0,0.0)(1,0.010737464442695086)(2,0.00010915287483787406)(3,4.269500398141168e-05)(4,0.0005910392990801233)(5,0.0009542708074993916)(6,0.0030700479675405032)(7,0.029543650049072924)};
\addlegendentry{Feature shortcoming}
\addplot[color=magenta,]
coordinates {(0,0.008928573690354824)(1,0.05744304184552883)(2,0.03174738570226962)(3,0.031203727240504714)(4,0.04450352271472679)(5,0.04296277944262461)(6,0.05730878857739812)(7,0.06559963771266453)};
\addlegendentry{Version}

\end{axis}
\end{tikzpicture}    
    \caption{The relative impact scores of different categories between February 2020 and September 2020}
    \label{fig:relimpact}
\end{figure*}
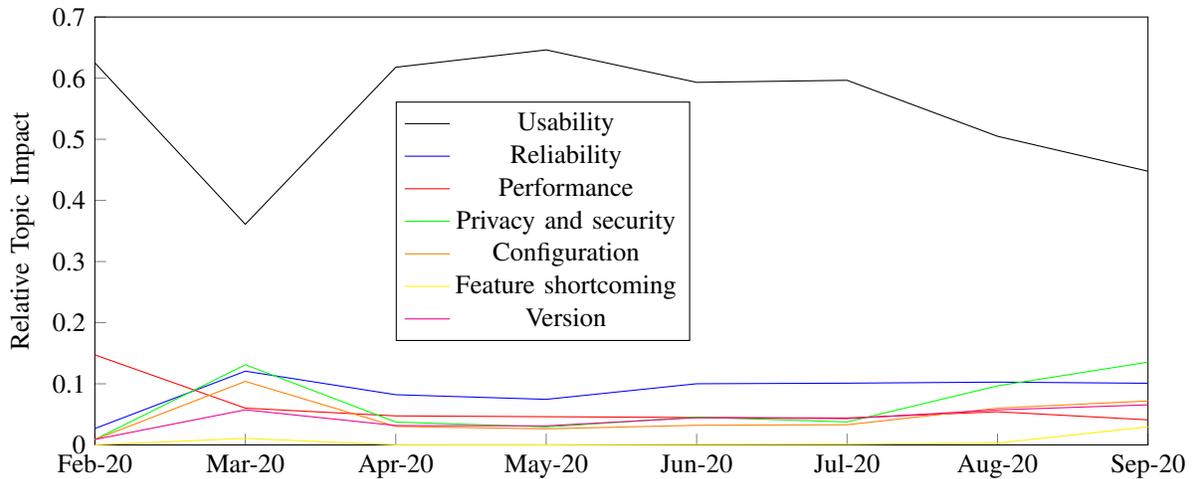

The absolute impact score trend and the relative impact score trend of different categories between the period February 2020 and September 2020 are shown in \fig\ref{fig:absimpact} and \fig\ref{fig:relimpact} respectively. \\
It shows that the absolute impact of Usability was significantly higher compared to other categories during the period of April-July. A possible reason is that user satisfactions are likely to be expressed immediately at the initial stages after the release (March in this context). Also, users might get used to the UX limitations within a few months, and negative UX reviews are likely to decline in number, too. 
The relative impact score plot in \fig\ref{fig:relimpact} shows the temporal evolution of topics under a particular category. Usability is again the dominating category for the whole period. The relative scores of the categories Reliability, Performance, and Version remain more or less stable over time. Interestingly, there is an indication of an increase in the trend of Privacy and security in the next two months.
\begin{figure*}[h]
\centering
\begin{tikzpicture}
\begin{axis}[
    scaled ticks=false, tick label style={/pgf/number format/fixed},
    ylabel={Absolute Topic Impact},
    x=2cm,
    xmin=0, xmax=7,
    ymin=0, ymax=95000,
    xtick={0, 1, 2, 3, 4, 5, 6, 7},
    xticklabels={Feb-20, Mar-20, Apr-20, May-20, Jun-20, Jul-20, Aug-20, Sep-20},
    ytick={0,5000,10000,15000,20000,25000,30000,35000,40000,45000,50000,55000,60000,65000,70000,75000,80000,85000,90000,95000},
    xtick pos=bottom,
    ytick pos=left,
    legend pos=north east,
    ymajorgrids=true,
    grid=none,
]
\addplot[color=black,]
coordinates {(0,1.2500014379620552)(1,244.3598127253354)(2,56308.630863203674)(3,93332.03008675948)(4,26586.250179078434)(5,17194.598425699398)(6,6387.4707652116185)(7,4760.58281943202)};
\addlegendentry{Usability}
\addplot[color=blue,]
coordinates {(0,0.053571442142128944)(1,81.70699408091605)(2,7487.8425128152585)(3,10772.561477211304)(4,4486.835630888119)(5,2912.5128797041248)(6,1300.0076315263284)(7,1072.4466834869236)};
\addlegendentry{Reliability}
\addplot[color=red,]
coordinates {(0,0.2946385703980923)(1,40.76511607132853)(2,4328.755369000137)(3,6668.732567746191)(4,2011.3813239531592)(5,1270.24479746446)(6,686.5305795138703)(7,436.4789954740554)};
\addlegendentry{Performance}
\addplot[color=green,]
coordinates {(0,0.017857147380709648)(1,88.82920171692966)(2,3416.067504116334)(3,4260.572029883972)(4,2041.437277683057)(5,1084.0718914791942)(6,1222.9272838048635)(7,1439.96610724926)};
\addlegendentry{Privacy and security}
\addplot[color=orange,]
coordinates {(0,0.017857147380709648)(1,70.37972011603415)(2,2751.978258866817)(3,3805.88488020096)(4,1441.2648042095825)(5,952.2747704768553)(6,760.8123753797263)(7,764.6336349966003)};
\addlegendentry{Configuration}
\addplot[color=yellow,]
coordinates {(0,0.0)(1,7.269263427704573)(2,9.948411318473518)(3,6.166354035027325)(4,26.492745541967448)(5,27.50876456778496)(6,38.8391768373549)(7,314.10808732174337)};
\addlegendentry{Feature shortcoming}
\addplot[color=magenta,]
coordinates {(0,0.017857147380709648)(1,38.88893932942301)(2,2893.5202276762566)(3,4506.691917891614)(4,1994.8259021649137)(5,1238.4880429925395)(6,725.0134842926635)(7,697.4553481610492)};
\addlegendentry{Version}
    
\end{axis}
\end{tikzpicture}    
    \caption{The absolute impact scores of different categories between February 2020 and September 2020}
    \label{fig:absimpact}
\end{figure*}
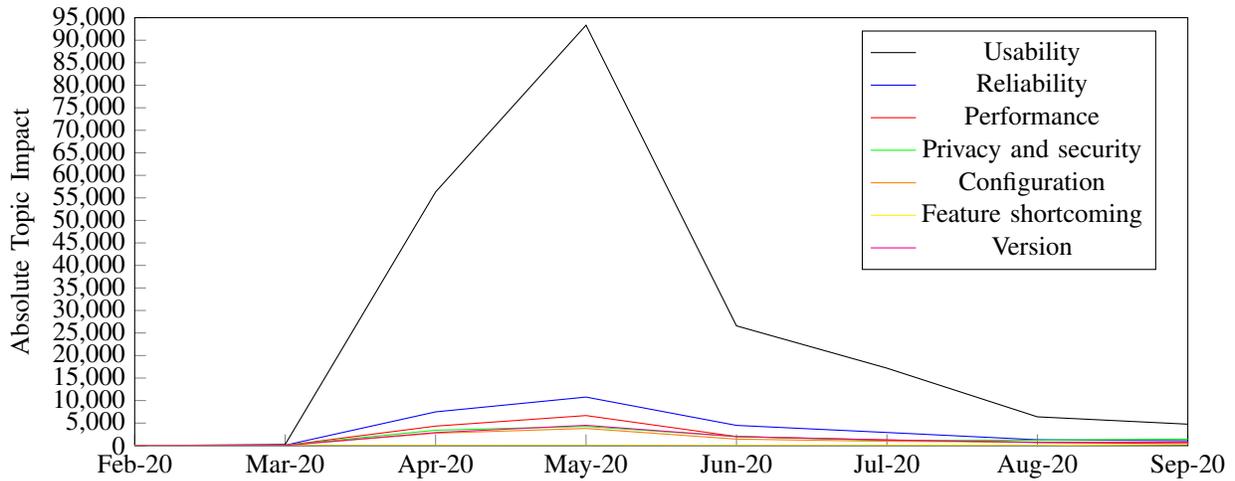

\subsection{User complaints and appraisals across topics (RQ3)}

\subsubsection{Motivation} User feedback can be associated with a number rating in the range of 1 to 5.  Typically, users give a high rating when they are pleased with the app, and they give a low rating when they find the app problematic. Star rating is the simplest metric to analyze the users' emotional expressions. Analyzing the star rating across different categories of the topics would help understand the degree to which users appreciate or dissent the app. This study outcome would help developers assign resources to address the issues belonging to the topics with a high ratio of complaints. 
\subsubsection{Approach}
For each of the seven categories, we calculate the proportion of appraising reviews and complaining reviews. We consider a review as an appraisal if the corresponding star rating is greater or equal to 3. Otherwise, we consider the review as a complaint.

\subsubsection{Results}
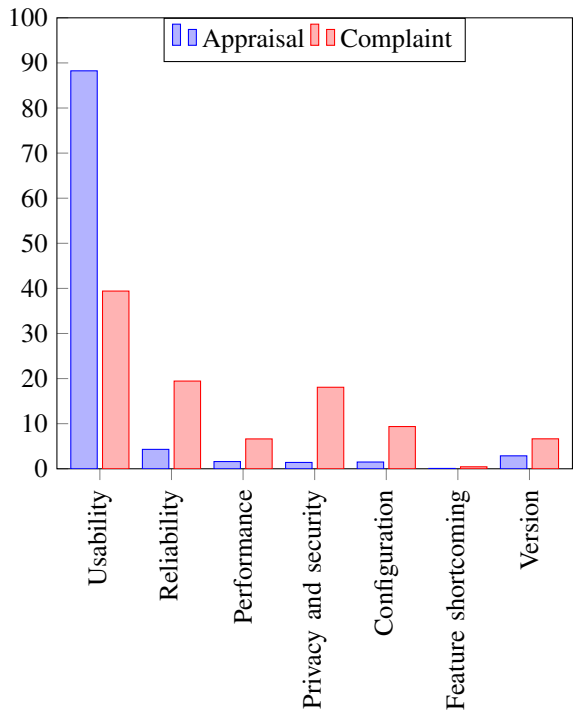
\begin{figure}[h]
\centering
\begin{tikzpicture}
\begin{axis}[
	    x tick label style={rotate=90,/pgf/number format/1000 sep=},
	    y=0.06cm,
	    ymin=0,ymax=100,
	    ytick={0,10,20,30,40,50,60,70,80,90,100},
	    xtick pos=bottom,
        ytick pos=left,
    	enlarge x limits=0.1,
    	legend style={at={(0.5,1)},
    	anchor=north,legend columns=-1},
    	ybar,
        symbolic x coords={Usability, Reliability, Performance, Privacy and security, Configuration, Feature shortcoming, Version},
        xtick={Usability, Reliability, Performance, Privacy and security, Configuration, Feature shortcoming, Version},
    ]
    \addplot coordinates {
        (Usability,88.2465657019625)
        (Reliability,4.303367928899665)
        (Performance,1.601220774062732)
        (Privacy and security,1.4050747131842904)
        (Configuration,1.497399235622633)
        (Feature shortcoming,0.07629837892074376)
        (Version,2.8700732673474296)
    };
    \addplot coordinates {
        (Usability,39.40674293638412)
        (Reliability,19.448777951118043)
        (Performance,6.621598197261224)
        (Privacy and security,18.077223088923557)
        (Configuration,9.364707921650199)
        (Feature shortcoming,0.43985092737042814)
        (Version,6.641098977292425)
    };\legend{Appraisal,Complaint}
\end{axis}
\end{tikzpicture}
    \caption{The star rating of topics across different categories}
    \label{fig:topicstarrating}
\end{figure}
\fig\ref{fig:topicstarrating} shows the proportions of appraisals and complaints for each topic category. As expected, most of the appreciative ratings are from the reviews under Usability. It was evident from the results of RQ1 in \tbl\ref{tab:challengespertopic} that most of the reviews in Usability are about positive UX. Therefore, 88\% of users reviewing UX gave high ratings. The other topic categories contain more complaints than appraisals, as reviews of these categories involve the app's failure to live up to the users' expectations.

\subsection{Evolution of user complaints and appraisals across the continents (RQ4)}

\subsubsection{Motivation} Conducting a comparative analysis of the star ratings of different continents can capture how the overall feeling of the users about the app varies from continent to continent. The outcome of this analysis, along with that of RQ2, will allow the developers of an app for a particular continent to understand the expectations of their potential users.

\subsubsection{Approach}
For the two classes of reviews, i.e., appraisal and complaint, we calculate the percentage of reviews belonging to each class for the seven continents and graphically show the results.
\subsubsection{Results} We observe that all continents except Australasia have more reviews about appraisals than complaints. For Australasia, the percentage of the two classes differ slightly. Asia has the highest portion of appreciative reviews among all continents. Again, the contact tracing app of India contributed to this high percentage.
\begin{figure}[t]
\centering
\begin{tikzpicture}
\begin{axis}[
	    x tick label style={rotate=90,/pgf/number format/1000 sep=},
	    y=0.06cm,
	    ymin=0,ymax=100,
	    ytick={0,10,20,30,40,50,60,70,80,90,100},
	    xtick pos=bottom,
        ytick pos=left,
    	enlarge x limits=0.1,
    	legend style={at={(0.5,1)},
    	anchor=north,legend columns=-1},
    	ybar,
        symbolic x coords={Europe,North America,Australasia,Asia,Middle East,Africa,Latin America},
        xtick={Europe,North America,Australasia,Asia,Middle East,Africa,Latin America},
    ]
    \addplot coordinates {
        (Europe,58.93019038984588)
        (North America,73.05757406372275)
        (Australasia,49.1107347025196)
        (Asia,88.80007823450794)
        (Middle East,53.94313649813458)
        (Africa,79.01023890784981)
        (Latin America,58.208955223880594)
    };
    \addplot coordinates {
        (Europe,41.06980961015413)
        (North America,26.94242593627725)
        (Australasia,50.88926529748042)
        (Asia,11.199921765492062)
        (Middle East,46.05686350186542)
        (Africa,20.98976109215017)
        (Latin America,41.791044776119406)
    };\legend{Appraisal,Complaint}
\end{axis}
\end{tikzpicture}
    \caption{The star rating of reviews across different continents}
    \label{fig:contstarrating}
\end{figure}
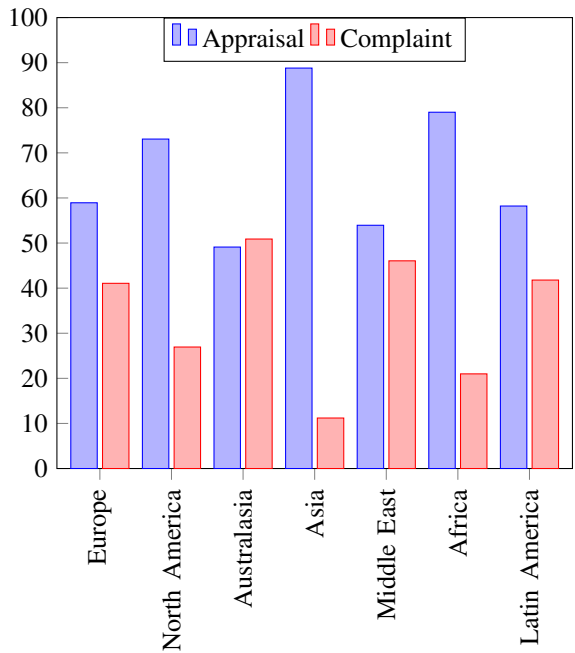

%% file: discussions.tex
\section{Discussions} \label{sec:discussion}

This study is expected to provide actionable insight to the developers of software/mobile applications as well as benefit the users. The implications of our findings are discussed below.

\subsection{Implications to the contact tracing app developers}
The major findings of this study are related to the user reactions expressing satisfaction about certain features of different apps and also their complaints about some issues or lack of features. Contact tracing application developers are the direct recipient of the insight obtained from this systematic study. They can better plan to address the issues in upcoming releases. They can be cautious while developing new features that users do not experience the same problems again. They also understand the differences in expectations of users or their experienced issues related to the surrounding environment of different geo-political regions.

\subsection{Implications to the emergency response app developers }
Contact tracing apps were developed hastily as an emergency response. These apps need to be developed without enough spending effort to design features, follow best programming practice, and test comprehensively as the developers have to meet a crying need. Many of the findings of this study would benefit the engineers of any app developed under similar circumstances. An example at hand is the current need for apps for the management of vaccination across the countries. Many issues found from the feedback of contact tracing apps are relevant here. Hence, the benefit of this study to the developers is beyond the immediate impact on contact tracing app providers.

\subsection{Implications to the users}
The most prevalent issues identified through this study are expected to attract the attention of developers, and they can prioritize their fixing plans accordingly. Consequently, there will be a quick release of more robust and user-friendly versions. The users will be relieved of their concerns and feel comfortable to use the apps when they find that their needs are addressed. This is likely to create an overall positive reputation that should increase the number of users significantly. The success of the attempts like contact tracing through apps depends on the size of a large user-base. Hence, this study is expected to contribute to widening the user-base and thus benefit the ultimate cause to a great extent. 

%% file: threats.tex
\section{Threats to Validity}\label{sec:threats}
We discuss the threats to validity of our studies following common guidelines
for empirical studies~\cite{Woh00}.

\bf{External Validity}. The primary external threat to validity is the generalizability of our findings. Our study revolves around the reviews of COVID-19 contact tracing apps. Therefore, the results of our research may not generalize to non-COVID domains. Moreover, our analysis involves reviews from February 2020 to September 2020. A drastic change in the review patterns could make our approach inapplicable to the new reviews in the coming months.

\bf{Internal Validity} The internal threats to validity concern experimental bias and errors while
conducting the analysis. The internal threats involving topic modeling are the performance degradation of LDA on short texts and selecting the optimal number of topics. LDA leverages word co-occurrence patterns to infer the latent topic distribution of documents. However, the lack of word co-occurrence information in short text causes a decrease in LDA performance. In our dataset, the LDA model fails to assign a particular topic to some short reviews. For example, the review "spy app" is assigned to multiple topics of more than one category with the same probability. Choosing the optimal number of topics is a complicated process. To address this threat, we use the four-stage topic coherence metric in \cite{roder2015exploring} to select the appropriate value of the number of topics. Moreover, LDA renders different results on the same corpus for multiple runs. Hence, we run LDA models ten times and consider the overall best number of topics that yields the desired score. We noticed a minimal difference in the coherence value for multiple runs on the optimal number of topics.
\\
\bf{Construct Validity} The threat of Construct Validity, in our context, stems from the possibility that labeling the resulting topics from the LDA might not reflect the reviews associated with the topics. To minimize this threat, the first two authors individually examined 15 reviews having the highest probability belonging to each topic to assign labels to that topic.
Then they discussed with the third and fourth authors to reach a consensus about the labels to ensure that the labels correctly represent the topics. To merge the topics into subcategories and categories, similarly, the first two authors did the initial assignment, and those were validated by detailed discussion involving all four authors. Since this is a standard practice in SE research, we believe it substantially reduces the threat to construct validity. 

%% file: conclusion.tex
\section{Conclusions} \label{sec:conclusion}

This paper presents a study on different topics the users discuss about the COVID-19 contact tracing apps across the continents by analyzing the reviews of the apps. It reveals specific problems users had to face, the concerns they have about the apps, and the positive/negative experience they share about the design and usage of the apps. We identified around 31 topics from the 35 apps under study from all over the world. Topics related to usability and performance of the apps are prevalent across all regions. Users often complained about the limitation of features and complex user interfaces in the apps, as well as the negative effect of such apps on their mobile batteries. Despite some negative experiences, many users considered the app as a utility tool to be wary of the potential sources of infection. Users also expressed concerns about the privacy implications of the app. Our study analyzes all these kinds of topics in a structured manner. The outcome of this study is expected to benefit the developers to address the issues in a planned way and release improved versions with informed decisions.